# Accepted Manuscript

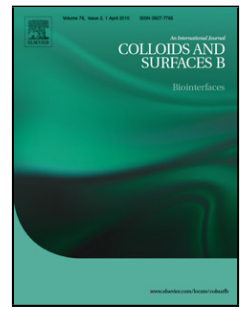

Title: Interaction of surfactant and protein at the O/W interface and its effect on colloidal and biological properties of polymeric nanocarriers

Authors: Teresa del Castillo-Santaella, José Manuel Peula-García, Julia Maldonado-Valderrama, Ana Belén Jódar-Reyes

Please cite this article as: del Castillo-Santaella T, Peula-García JM, Maldonado-Valderrama J, Jódar-Reyes AB, Interaction of surfactant and protein at the O/W interface and its effect on colloidal and biological properties of polymeric nanocarriers, *Colloids and Surfaces B: Biointerfaces* (2018), https://doi.org/10.1016/j.colsurfb.2018.09.072

# Interaction of surfactant and protein at the O/W interface and its effect on colloidal and biological properties of polymeric nanocarriers

*Teresa del Castillo-Santaella[a], José Manuel Peula-García[a,b], Julia Maldonado-Valderrama[a,c], and Ana Belén Jódar-Reyes[a,c,*]*

[a]Biocolloid and Fluid Physics Group, Department of Applied Physics, University of Granada, 18071 Granada, Spain.

[b]Department of Applied Physics II, University of Málaga, 29071 Málaga, Spain.

[c]Excellence Research Unit "Modeling Nature" (MNat), University of Granada, Granada, Spain.

*corresponding author: Ana Belén Jódar Reyes (ajodar@ugr.es). Department of Applied Physics; Faculty of Sciences, *Campus de Fuentenueva, sn. 18071, Granada, Spain, Tel: +34 958248530;Fax.+34 958243214*

E-mail addresses: tdelcastillo@ugr.es; jmpeula@uma.es; julia@ugr.es;

**GRAPHICAL ABSTRACT**

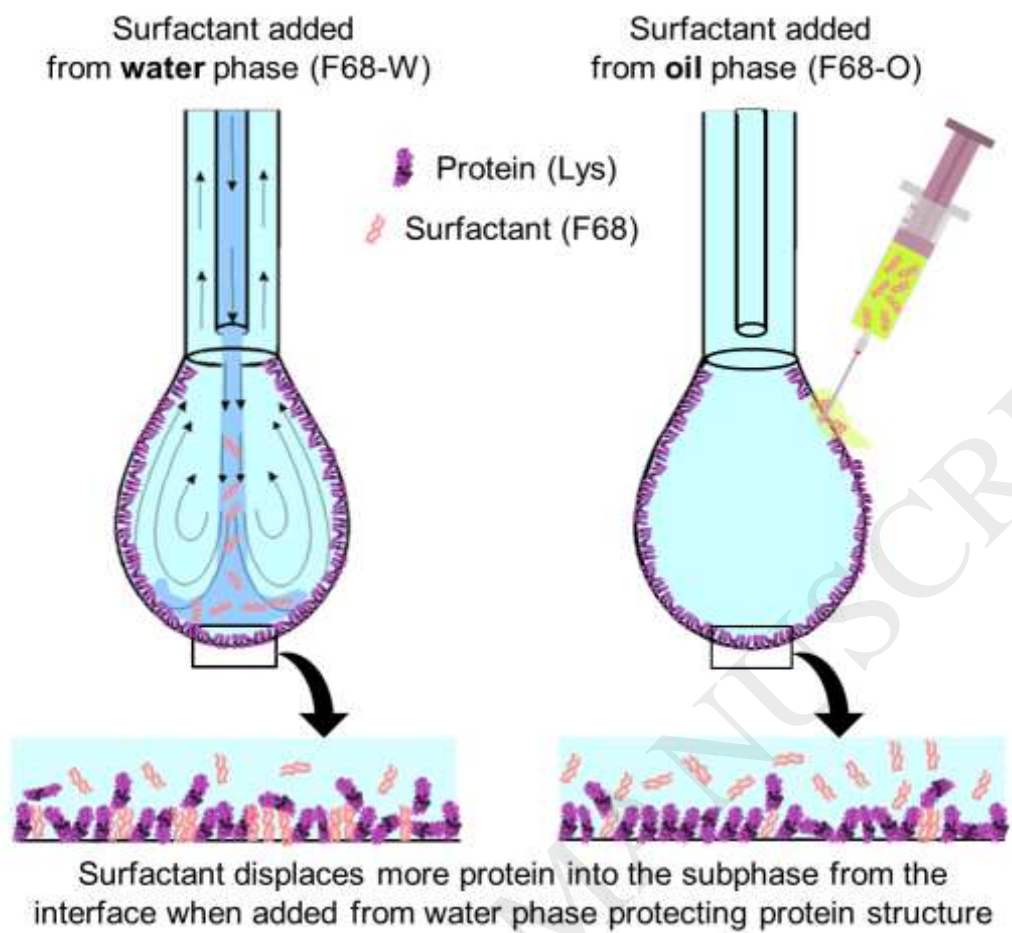

HIGHLIGHTS

- *Interfacial and colloidal properties underlying synthesis of NPs by W/O/W technique*
- *Surfactant (F68) added from water improves colloidal and biological NP properties*
- *An in vitro model to mimic the first step of the W/O/W NP synthesis is developed*
- *F68 added from water displaces more protein from interface protecting its structure*

# ABSTRACT

*Hypothesis*

The use of polymer-based surfactants in the double-emulsion (water/oil/water, W/O/W) solvent-evaporation technique is becoming a widespread strategy for preparing biocompatible and biodegradable polymeric nanoparticles (NPs) loaded with biomolecules of interest in biomedicine, or biotechnology. This approach enhances the stability of the NPs, reduces their size and recognition by the mononuclear phagocytic system, and protects the encapsulated biomolecule against losing biological activity. Different protocols to add the surfactant during the synthesis lead to different NP colloidal properties and biological activity.

*Experiments*

We develop an *in vitro* model to mimic the first step of the W/O/W NP synthesis method, which enables us to analyze the surfactant-biomolecule interaction at the O/W interface. We compare the interfacial properties when the surfactant is added from the aqueous or the organic phase, and the effect of pH of the biomolecule solution. We work with a widely used biocompatible surfactant (Pluronic F68), and lysozyme, reported as a protein model.

*Findings*

The surfactant, when added from the water phase, displaces the protein from the interface, hence protecting the biomolecule. This could explain the improved colloidal stability of NPs, and the higher biological activity of the lysozyme released from nanoparticles found with the counterpart preparation.

**ABBREVIATIONS** (Alphabetical order)**:**

**ADSA** (axisymmetric drop-shape analysis)

**DLS** (Dynamic Light Scattering)

**DMC** (dichloromethane)

**E** (dilatational modulus)

**EA** (ethyl acetate)

**EE** (protein-encapsulation efficiency)

**F68-O** (procedure in which the Pluronic F68 was dissolved in the organic phase)

**F68-O-Lys NPs** (lysozyme-loaded nanoparticles resulting from the F68-O method)

**F68-O-Lys5.5**, **F68-O-Lys9**, **F68-O-Lys12** (lysozyme-loaded nanoparticles resulting from the F68-O method, when the pH of the lysozyme solution is: 5.5, 9.0 or 12.0)

**F68-W** (procedure in which the Pluronic F68 was dissolved in the aqueous phase)

**F68-W-Lys NPs** (lysozyme-loaded nanoparticles resulting from the F68-W method)

**F68-W-Lys5.5**, **F68-W-Lys9**, **F68-W-Lys12** (lysozyme-loaded nanoparticles resulting from the F68-W method, when the pH of the lysozyme solution is: 5.5, 9.0 or 12.0)

**i.e.p.** (isoelectric point)

**$M_E$** (final encapsulated amount of lysozyme)

**$M_F$** (total mass of lysozyme in the aqueous supernatant)

**$M_I$** (initial total mass of lysozyme)

**$M_{polymer}$** (mass of PLGA in the formulation)

**MPS** (mononuclear phagocytic system)

**NPs** (Nanoparticles)

**NTA** (Nanoparticle Tracking Analysis)

**O/W** (Oil/Water)

**PB** (phosphate buffer)

**PDI** (polydispersity index)

**PEO** (Poly(ethylene oxide))

**PL** (final protein loading)

**PLGA** (Poly(lactide-co-glycolide) Acid)

**SEM** (scanning electron microscopy)

**SESD** (spontaneous emulsification solvent diffusion)

**STEM** (scanning transmission electron microscopy)

**W/O/W** (Water/Oil/Water)

**μ-average** (average electrophoretic mobility).

# 1. INTRODUCTION

Nanometer-scale biocompatible and biodegradable polymeric particles, such as those formed with polylactide glycolic acid (PLGA), are designed and optimized to carry a wide variety of biomolecules. They have been widely studied for use as drug-delivery vehicles for long-term sustained-release preparations[1,2].

Several methods are available for preparing PLGA NPs and for incorporating biomolecules into them depending on the biomolecule characteristics, the desired delivery path, and the release profile. The spontaneous emulsification solvent diffusion (SESD) method is the basis for different methods of preparing polymeric NPs. Nanosized particles can be synthesized by pouring a PLGA organic solution into an aqueous phase (or surfactant solution) with mechanical stirring and finally a solvent-evaporation process. For the preparation of NPs loaded with biomolecules, the SESD method is modified, and a double-emulsion (water/oil/water, W/O/W) solvent-evaporation technique is used [3]. In this case, a biomolecule water (or buffered) solution is added onto the organic polymeric solution and mixed by mechanical energy. This first water/oil (W/O) emulsion is immediately poured into the second polar phase.

The addition of stabilizers during the preparation method, such as poly(ethylene oxide) (PEO) surfactants, is a promising way to protect the biomolecule from losing activity during its encapsulation, storage, delivery, and release [4–9]. The use of the polymeric surfactant Pluronic F68 also reduces the size of the NPs, and enhances their stability. In addition, the recognition of the nanocarriers by the mononuclear phagocytic system (MPS) is reduced.

In a previous work [10], we developed and optimized two different formulation methods for protein-loaded NPs (PLGA colloidal particles) based on the double-emulsion W/O/W solvent-evaporation technique. They differed mainly in the phase the surfactant (Pluronic F68) was added from. In both cases we obtained hard spherical NPs, but with different colloidal properties (size distribution, electrokinetic charge, colloidal stability) hence strongly influencing cellular uptake. In particular, we obtained improved *in vitro* biological activity of the released protein, and better release pattern when the surfactant was added from an aqueous phase [10]. The protein used was lysozyme, as it is considered to be a model for proteins having potential therapeutic applications (e.g. bone morphogenetic proteins) [11,12].

Accordingly, the aim is this work is to evaluate in detail how the solvent used for the surfactant and the conditions of the protein solution can affect ultimately the biological activity and colloidal stability of the NPs, by determining the protein/surfactant interactions and the interfacial composition. To this end, we analyze the properties of the protein at the interface as a function of: a) the procedure to add the surfactant (from the water or oil phase), and b) the conditions of the protein solution (pH). The interfacial results importantly correlate with the properties of the colloidal systems synthesized following the corresponding conditions and explain the different biological activity encountered depending on the method of preparation used [10].

## 2. MATERIALS AND METHODS

### 2.1. Formulation of the nanoparticles

Poly(lactide-co-glycolide) Acid (PLGA 50:50) ([$C_2$ $H_2$ $O_2$]x [$C_3$ $H_4$ $O_2$]y) x=50, y=50 (Resomer® 503H), 32-44 kDa was used as the polymer, and polymeric surfactant Pluronic®F68 (Poloxamer 188) (Sigma-Aldrich-P7061) served as the emulsifier. The structure, based on a poly(ethylene oxide)-block-poly(propylene oxide)-block-poly(ethylene oxide), is expressed as PEOa-PPOb-PEOa with a=75 and b=30. Lysozyme from chicken egg white (Sigma-L7651) was used as a hydrophilic protein. This is a small globular with 129 amino acids, a molecular weight of 14300 g/mol, and an isoelectric point (i.e.p.) of 11.35 [13].

Ultrapure water, passed through a Milli-Q water-purification system (0.054 mS), was used to prepare the buffer solutions. All glassware was washed with 10% Micro-90 cleaning solution and exhaustively rinsed with tap water, isopropanol, deionized water, and ultrapure water (in that sequence). All other chemicals were of analytical grade and used as received.

The two different formulation methods used were based on those we developed in a previous work [10]. Here, we use the terms F68-O and F68-W to designate the procedure in which the Pluronic F68 was dissolved in the organic (O) and aqueous phase (W), respectively. For both methods, the primary water/oil (W/O) emulsion has been prepared using three different pH conditions for the lysozyme buffered solution: 5.5 (water), 9 (boric acid 0.1 M), and 12 (di-sodium phosphate 0.03 M). Lysozyme presents the highest charge at pH 5.5, falling to 75% of that value at pH 9.0, both being positive. At pH 12.0 the charge turns negative with an absolute value of 38% of its value at pH 5.5.[14]

Briefly, in the F68-O method, 25 mg of PLGA, and 15 mg of Pluronic F68 were dissolved in 660 µL of dichloromethane (DMC), and vortexed. Then, 330 µL of acetone were added, and vortexed. Next, 100 µL of an aqueous buffered solution with lysozyme (5 mg/mL) were added dropwise while vortexing for 30 sec. This primary W/O emulsion was immediately poured into a glass containing 12.5 mL of ethanol under magnetic stirring, and 12.5 mL of MilliQ water were added. After 10 min of magnetic stirring, the organic solvents were rapidly extracted by evaporation under vacuum until the sample reached a final volume of 10 mL.

In the F68-W method, 100 mg of PLGA were dissolved in a tube containing 1 mL of ethyl acetate (EA), and vortexed. To prepare the primary emulsion, 40 µL of a buffered solution with lysozyme (20 mg/mL) were added and immediately sonicated (Branson Ultrasonics 450 Analog Sonifier), fixing the *duty cycle* dial at 20% and the *output control* dial at 4, for 1 min with the tube surrounded by ice. This primary W/O emulsion was poured into a plastic tube containing 2 mL of a buffered solution (pH 12.0) of F68 at 1 mg/mL, and vortexed for 30 sec. Then, the tube surrounded by ice was sonicated again at the maximum amplitude for the micro tip (*output control* 7), for 1 min. This second W/O/W emulsion was poured into a glass containing 10 mL of the buffered F68 solution and kept under magnetic stirring for 2 min. The organic solvent was then rapidly extracted by evaporation under vacuum to a final volume of 8 mL.

For clarity, the lysozyme-loaded nanoparticles resulting from the two methods described above are designated as F68-O-Lys and F68-W-Lys NPs. When the pH of the lysozyme solution is specified (pH 5.5, 9.0 or 12.0) we will refer to the NPs as F68-O-Lys5.5, F68-O-Lys9, F68-O-Lys12 (for the F68-O method), and F68-W-Lys5.5, F68-W-Lys9, F68-W-Lys12 (for the F68-W method).

*Cleaning and storage*

After the organic solvent evaporation, the sample was centrifuged during 10 min at 20ºC at 14000 and 12000 rpm for F68-O and F68-W methods, respectively. The supernatant was filtered using 100 nm filters for measuring the free non-encapsulated protein. The pellet was then resuspended in PB up to a final volume of 4 mL and kept refrigerated at 4ºC.

*Protein loading and encapsulation efficiency*

The initial protein loading was optimized for the nanoparticle formulation, preserving the final colloidal stability after the evaporation step and being different for each nanosystem. The protein-encapsulation efficiency (EE) was determined from the initial total mass of lysozyme ($M_I$), and the total mass of lysozyme in the aqueous supernatant ($M_F$), which corresponded to the free non-encapsulated protein, and was tested by the bicinchoninic acid assay (BCA, Sigma-Aldrich).

$$EE = \frac{M_I - M_F}{M_I} \times 100$$

For the final protein loading (PL), the mass of PLGA in the formulation ($M_{polymer}$) was also taken into account:

$$PL = \frac{M_I - M_F}{M_{polymer}} \times 100$$

## 2.2. Characterization of the nanoparticles

*Nanoparticle size and electrokinetic mobility*

The hydrodynamic diameter and electrophoretic mobility of the NPs were determined by using a Zetasizer NanoZeta ZS device (Malvern Instrument Ltd, U.K.) working at 25ºC with a He-Ne laser of 633 nm, and a 173ºscattering angle. Each data point was taken as an average over three independent sample measurements. Dynamic Light Scattering (DLS) was used to determine the average hydrodynamic diameter (Z-average or cumulant mean), and the polydispersity index (PDI) of the samples through a cumulant analysis of the data, which is applicable for narrow monomodal size distributions[15]. In other cases, an algorithm included in the Zetasizer software (General Purpose) was applied to calculate the intensity size distribution.

For each sample, the electrophoretic mobility distribution, and the average electrophoretic mobility (μ-average) was obtained by the technique of Laser Doppler Electrophoresis.

To gain complementary information on the hydrodynamic size distribution of the NPs, we used Nanoparticle Tracking Analysis (NTA) with a NanoSight LM10-HS(GB) FT14 (NanoSight, Amesbury, U. K.). The particle concentration as a function of the diameter (size distribution) was calculated as an average of at least three independent size distributions. All samples were measured at 25ºC for 60 s with manual gain, brightness, shutter, and threshold adjustments.

*Colloidal stability vs. time*

The average hydrodynamic diameter and the polydispersity index (PDI) by DLS, and the size distribution by NTA of each NP system in phosphate buffer (PB) were determined one month after the synthesis, and compared with the previous measurements to evaluate the colloidal stability of these samples under storage at 4ºC.

### 2.3. Interfacial characterization of the first water-in-oil emulsion

The surface tension and dilatational rheology measurements were made in the OCTOPUS. This is a Pendant Drop Surface Film Balance equipped with a subphase multi-exchange device (patent submitted P201001588) described in detail elsewhere [16–18]. The surface tension was recorded at a constant interfacial area (20 mm$^2$). The dilatational rheology of the surface layer was carried out at amplitude values of < 5%, in order to avoid excessive surface perturbation, while the measurement frequency (ν) was set to 1 Hz. The surface tension was calculated with DINATEN® software, based on axisymmetric drop-shape analysis (ADSA), and the dilatational modulus (E) of the interfacial layer was determined from image analysis with the program CONTACTO®. This device enabled us to study a sequential static adsorption of amphiphilic molecules on a single droplet, calculating the surface tension and the dilatational elasticity of the interfacial layers formed, as described in detail elsewhere [16–19].

Lysozyme (4 mg/ml) was dissolved in water (pH 5.5), and H$_3$BO$_3$ buffer (pH 9.0, pH 12.0), whereas Pluronic F68 (15mg/ml) was dissolved in water or H$_3$BO$_3$ buffer (F68-W) or in chloroform (F68-O). Then, we design a sequential-adsorption experiment, mimicking the different processes used for nanoparticle preparation where the oil-water interface is represented by an air-water interface. The sequential-adsorption experiment consists of 3 steps. First, the lysozyme adsorbs at the air-water surface. Then, we add the Pluronic F68 either from the aqueous or from the organic phase (F68-W or F68-O, respectively), hence mimicking the different preparation methods. F68-W was added by subphase exchange of the bulk solution and F68-O was added by spreading onto the protein monolayer. A final step consisting of depletion of bulk material allows us evaluation of interfacial interactions and stability of interfacial layer [18].

The surface tension was recorded *in situ* throughout the three steps, and the dilatational elastic modulus of the interfacial layer was computed at the end of each step. The surface tension of the air-water interface was measured before each experiment to confirm the absence of surface active contaminants, yielding values of 72.5±0.5 mNm$^{-1}$ at 20°C. The reproducibility of the experiments was tested by performing at least three replicate measurements.

## 3.     RESULTS AND DISCUSSION

### 3.1 Nanoparticle synthesis

Several factors influence protein encapsulation inside PLGA nanoparticles produced by a double emulsion process. These factors need to be optimized to: a) design colloidally stable nanosystems; b) reach the best encapsulation efficiency of the bioactive protein; and c) find an adequate release pattern [3,4]. As it has been previously commented, the use of Pluronics in the formulation favors the optimization of these factors [5,7–10]. Moreover, we believe that the interaction between surfactant and protein could be determinant in the protection of the protein mainly during the first emulsion step. We compared two synthesis protocols developed in a previous work [10], in which the surfactant was added in the first emulsion from the organic phase (F68-O) or from the aqueous phase (F68-W). Additionally, we investigated how the electrical state of the lysozyme affects the interaction with the polymer and surfactant by changing the pH of the protein solution. Table 1 shows the encapsulation data corresponding to such conditions. The first notable result is the limiting effect of initial protein loading in preserving the colloidal stability of the first nanoemulsion. Thus, the initial protein-loading data shown in Table 1 correspond to the maximum values that led to colloidally stable nanosystems. The F68-O method with lysozyme solutions at pH 9.0 or in water (pH 5.5) gave rise to unstable samples. This occurred even by reducing the amount of protein below advisable values for effective bioactivity in encapsulated therapeutic proteins[7]. In both F68-O-Lys12 and F68-W-Lys12 systems, with the lysozyme solution at pH 12.0, the presence of Pluronic F68 in the first emulsion step led to a reduction of the EE in comparison with EE values in absence of surfactants. This is a consequence of altering the hydrophobic interaction between the protein and the polymer [5,10,20]. Despite a lower EE value for F68-O-Lys12 system, its PL value proved higher, perhaps related to a lower surfactant-protein interaction for the system in which surfactant comes from the oil phase. On the other hand, when we compared the different pH conditions for the F68-W-Lys NPs, we found a decrease in the EE by increasing the pH of the lysozyme solution. Although the protein-PLGA interaction shows an major hydrophobic contribution, it can be strongly influenced by the electrostatics [21,22]. As mentioned above, the electric charge of the lysozyme changes with pH. This strongly affects the protein-polymer interaction, and consequently the EE of lysozyme within the PLGA nanostructures [5]. When the pH is below the i.e.p. of the lysozyme (11.35), the attractive electrostatic interaction between negative terminal acid residues of PLGA and positive protein molecules enhances the protein encapsulation process, which affects the final protein loading [7,12]. The charge cancelation during the first step of the synthesis modulates the nanoparticle formation, inducing aggregation and influencing the nanosphere size distribution [23]. This could be the reason for the unstable systems obtained with lysozyme solutions at pH 9.0 or 5.5 using F68-O. However, these systems were stable using F68-W as will be explained below.

| | pH | $M_{polymer}$(mg) | $M_{F68}$(mg) | $M_I$(mg) | EE % | $M_E$ (mg) | PL % |
|---|---|---|---|---|---|---|---|
| **F68-O-Lys** | 12.0 | 25 | 15 | 0.4 | 62.5 | 0.25 | 1 |
| **F68-W-Lys** | 5.5 | 100 | 2 | 0.8 | 98.0 | 0.78 | 0.78 |
| | 9.0 | | | | 88.6 | 0.71 | 0.71 |
| | 12.0 | | | | 73.1 | 0.58 | 0.58 |

**Table 1**: *Formulation conditions and protein encapsulation results. $M_{polymer}$, $M_{F68}$ and $M_I$ are the initial amount of PLGA, Pluronic F68 and lysozyme, respectively; **EE** is the encapsulation efficiency; $M_E$ is the final encapsulated amount of lysozyme; **PL** is the final protein loading rate in w:w.*

Above the i.e.p. of the lysozyme, polymer and protein are negatively charged, and the corresponding electrostatic repulsion reduces the encapsulation. In this case, the electrostatic repulsion between protein and polymer beside the reduction of the hydrophobic protein-polymer interaction due to the presence of the Pluronic could explain the lowest PL value among all, F68-O and F68-W, nanosystems.

### 3.2. Nanoparticle size, electrokinetic mobility, and colloidal stability

Particle size strongly affects the biodistribution, delivery, and action mechanism of the transported protein. Micrometric particles are usually employed for a local supply and reduce the action of the phagocytic system [24]. Nevertheless, nanometric systems present higher stability and reactivity, provide a systemic distribution, and permit extra- and intracellular actions [25].

Table 2 lists the main colloidal properties of protein-loaded particles found with the two preparation methods at the different pHs for the lysozyme solution. As commented above, the F68-O method with lysozyme solutions at pH 9.0 and 5.5 gave rise to unstable samples, and macroscopic aggregates, and thus the corresponding data are not provided in this table.

| System | Diameter-Z-average (nm) | PDI | μ-average (μmcm/Vs) |
|---|---|---|---|
| F68-O-Lys12 | 212 ± 2 | 0.168 ± 0.005 | -5.11 ± 0.03 |
| F68-O-Lys9 | | Unstable | |
| F68-O-Lys5.5 | | | |
| F68-W-Lys12 | 293 ± 4 | 0.169 ± 0.016 | -4.212 ± 0.013 |
| F68-W-Lys9 | 175.2 ± 1.3 | 0.081 ± 0.011 | -5.15 ± 0.04 |
| F68-W-Lys5.5 | 570 ± 20 | 0.21 ± 0.03 | -3.30 ± 0.08 |

**Table 2**: *Colloidal properties of PLGA NPs from different synthesis methods at different lysozyme solution conditions. The properties were measured in phosphate buffer (pH 7.0). The errors correspond to the standard deviations of three independent sample measurements. The average hydrodynamic diameter (Z-average or cumulative mean) and the polydispersity index (PDI) were determined by DLS.*

The hydrodynamic diameter distributions of the particles were first determined by DLS (Figure 1A). The average hydrodynamic diameter (Z-average) and the polydispersity index (PDI) were calculated through a cumulative analysis of the data, which is applicable for narrow monomodal size distributions [15]. As reported previously [10], SEM and STEM micrographs of systems prepared with the same methods and with the lysozyme solution at pH 12.0 indicated that such an approximation could be assumed for particles from the F68-O, but not from the F68-W. Thus, the intensity size distributions (Figure 1A) of the different systems were obtained to gather better information.

Figure 1A indicates the presence of particles larger than 500 nm in the F68-W-Lys12 system, which does not correlate with the SEM micrographs [10]. Thus, a different technique (NTA) was used to gain more reliable information on the size distribution of such systems (Figure 1B). With NTA, the size distributions were consistent with the SEM images of the samples prepared using both methods (F68-W-Lys12 and F68-O-Lys12), reflecting that the protocol chosen has an effect on the size of the system [10]. Thus, a clear narrow peak in the size distribution was confirmed for F68-O-Lys12 NPs, with the highest particle concentration at 175 nm. For F68-W-Lys12 NPs, a wider distribution was observed, with the maximum particle concentration at 214 nm, but with a major contribution of particles of sizes up to 400 nm.

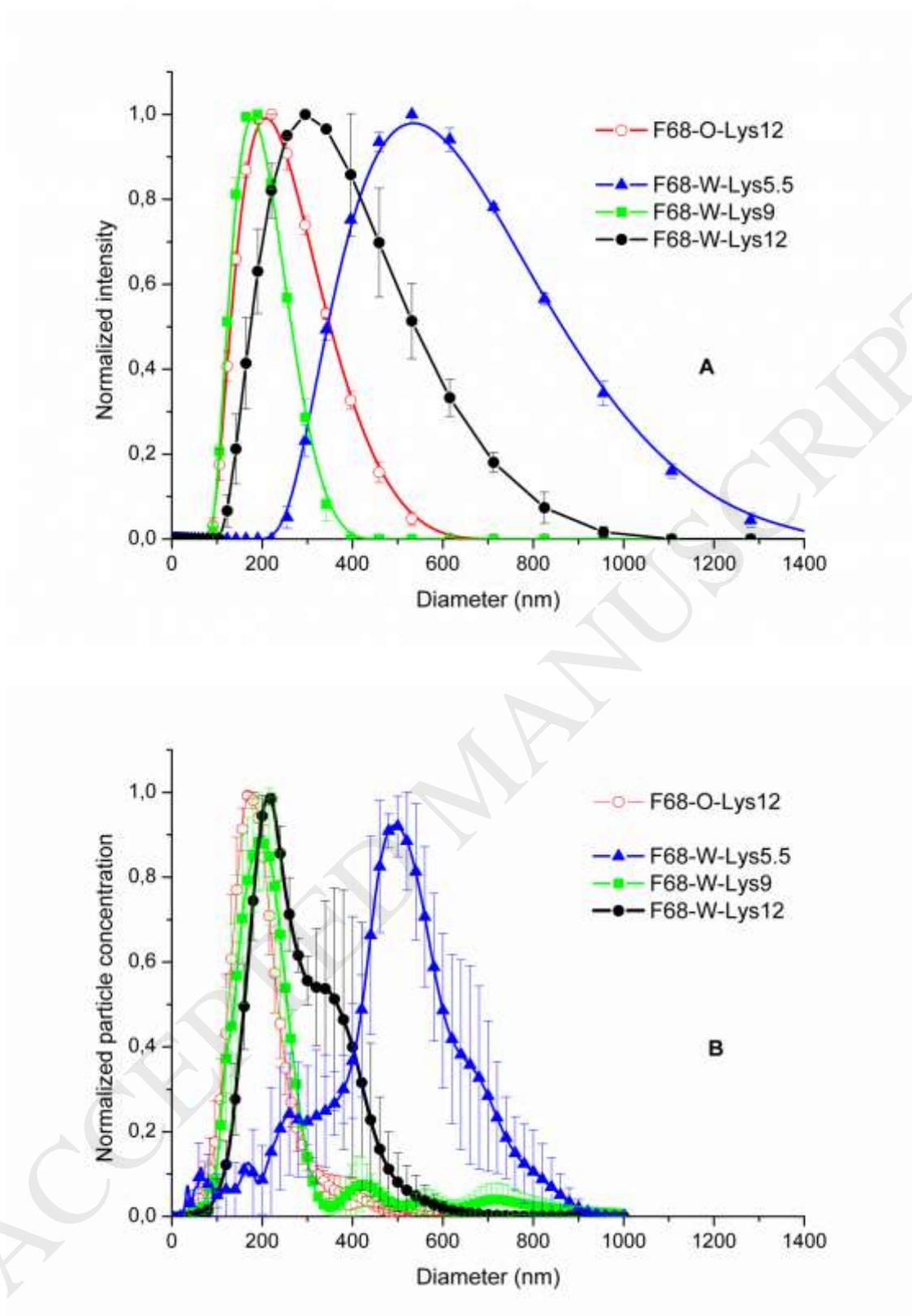

**Figure 1**: *Hydrodynamic diameter distribution of lysozyme-loaded PLGA particles from the F68-O and F68-W methods measured at pH 7.0 (phosphate buffer) (**A**) by DLS and (**B**) by NTA.*

When comparing the systems prepared with the same method, but with lysozyme solutions at different pHs, we found pronounced differences. F68-O-Lys5.5 and F68-O-Lys9 systems aggregated during the first W/O emulsion preparation. Conversely, stable particles were formed in all cases with F68-W. However, the F68-W-Lys5.5 system presented high polydispersity, with the maximum particle concentration at 490 nm, and notable contributions of between 600 and 700 nm. For F68-W-Lys12 NPs, as mentioned above, the distribution was wide, but sizes remained below 400 nm. A narrow peak with a maximum at 200 nm was observed for F68-W-Lys9 NPs, and much lower concentrations of particles with larger sizes were also detected (Figure 1B).

In summary, particles below 500 nm were found in F68-O-Lys12, and in F68-W-Lys9/Lys12 systems. NP sizes below 500 nm are necessary for cell internalization and rapid distribution after parenteral administration in order to reach different tissues through different biological barriers. These systems also maintained their size while under storage at 4ºC for at least 1 month (data not shown). However, particles in the range from 200 to 800 nm were detected in the F68-W-Lys5.5 system, and aggregates appeared after 1 month under storage at 4ºC.

The electrokinetic charge of the NPs was analyzed by electrophoretic mobility. With the aim of comparison, all the samples were measured at pH 7.0 (phosphate buffer). Figure 2 shows the electrophoretic mobility distributions and the corresponding average electrophoretic mobilities ($\mu$-average) are displayed in Table 2.

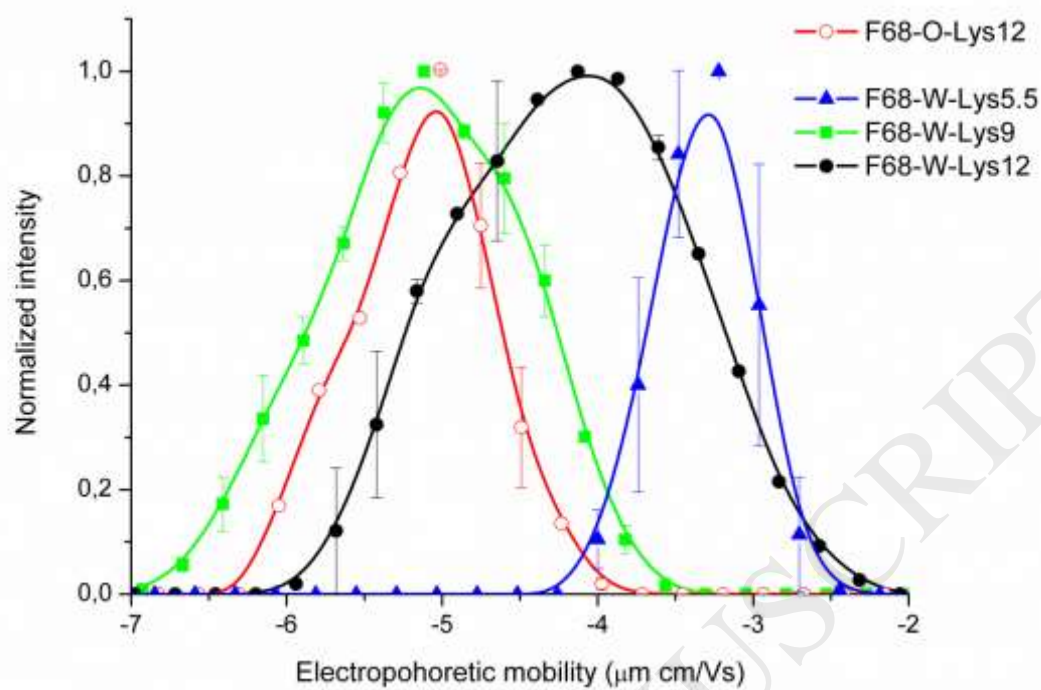

***Figure 2****: Electrophoretic mobility distribution of lysozyme-loaded PLGA particles from the F68-O and F68-W methods measured at pH 7.0 (phosphate buffer).*

All systems were negatively charged due to the carboxylic groups of the PLGA polymer. We previously checked that the use of Pluronic F68 in the F68-O clearly made the NPs less negative [10], indicating the presence of F68 at the NP surface, as also confirmed by NMR [10].

Lysozyme solution at pH 12.0 provided even less negative charged NPs indicating the presence of some protein (whose net charge is positive at pH 7.0) near or at the NP surface. F68-W provided less negative charge to the NPs indicative of more F68 and/or more lysozyme at the surface for F68-W-Lys12 NPs.

When comparing the different pH conditions of the lysozyme solution for F68-W, we observed that the F68-W-Lys9 NPs presented the more negative mobility (Table 2). The corresponding electrophoretic mobility distribution was similar to that of NPs resulting from the same method but without lysozyme[10], meaning that lysozyme was not present at or near the surface of the NPs. The lowest mobility in absolute value was found for F68-W-Lys5.5 NPs.

The electric charge of PLGA acid end groups and the Pluronic molecules located on the surface of the NPs confer a combined electrostatic and steric colloidal stability mechanism, as previously described [5,26]. This combined mechanism can explain that the NPs kept their size under storage at 4ºC at least for 1 month (data not shown) in all cases except for the F68-W-Lys5.5 system. This exception is consistent with the lowest electrophoretic charge of these NPs, as mentioned above.

### 3.3. Interfacial characterization of the first water (protein solution)-in- oil emulsion

The main differences in the synthesis methods under study were: a) how Pluronic F68 is added: in the aqueous phase (F68-W) or in the organic phase (F68-O); and b) the pH of the lysozyme solution (pH 5.5, 9.0 or 12.0). These differences could affect the composition of the O/W interface and, consequently, the colloidal properties of the NPs (Table 2).Thus, we designed a sequential adsorption process that mimics the different synthesis procedures and provides information on how the Pluronic F68 and lysozyme are organized on the interface of the first water (lysozyme solution)-in-oil emulsion. The OCTOPUS allows these procedures to be reproduced by adding the Pluronic F68 from the aqueous bulk solution or from the organic phase on a lysozyme-adsorbed surface layer [18].

Proteins behave differently when they are dissolved in aqueous medium or are between two immiscible liquids. In the first case, they show hydrophilic residues and hide hydrophobic residues towards the internal part of the protein. However, in the second case (i.e. air-water interface) proteins are adsorbed and suffer a partial denaturation to acquire a new conformation. The velocity and characteristics of the adsorbed protein layer depends on the thermodynamic stability, flexibility, amphipathicity, molecular size, and protein charge [27–30].

### Lysozyme-adsorbed layer at the air-water interface

The first step was to analyze the adsorption of lysozyme at the air-water interface, and the effect of the protein solution (pH) conditions on the resulting adsorbed layer.

Lysozyme is easily soluble in water but presents a rigid protein structure owed to strong intramolecular bonds (four disulfide bonds)[30,31]. Owing to this rigidity, most authors use denaturating agents to study lysozyme at interfaces [32,33]. However, in the present work, we sought to reproduce the conditions used in the synthesis of the nanoparticles described above and lysozyme was used directly, without prior denaturing.

In Table 3, we show the interfacial tension and the dilatational moduli attained after 1 h of adsorption of native lysozyme onto the air-water interface under different pH conditions but the same concentration (4 mg/ml). Considering the experimental errors, we found that the interfacial tension $\gamma_{ADS}$ was similar at the three pHs, only proving a slightly lower value at pH 12,0 (closer to the isoelectric point). The dilatational modulus of the adsorbed layer is very high for all three pHs owing to the rigidity of the protein due to its four disulfide bonds and in agreement with literature [30], conferring low mobility to the internal structure of lysozyme. The protein displayed the lowest dilatational modulus close to the isoelectric point, at pH 12.0, and the highest dilatational modulus when the lysozyme was more positively charged, at pH5.5. The dilatational modulus is affected by both intermolecular and intramolecular bonds formed at the interface. Hence, this tendency reflects the high rigidity of the protein, and the electrostatic repulsion at the interfacial layer resulting in electrosteric repulsion (see Table 3).The high values of the dilatational modulus for lysozyme adsorbed layers contrast with other flexible proteins, such as β-casein, which are highly surface active but form loose monolayers with lower dilational modulus [16]. Other globular proteins (ovoalbumin, β-lactoglobulin, human serum albumin) form rigid monolayers owing to a rigid internal structure with disulfide bonds[34].

To analyze the stability of the lysozyme-adsorbed layer at the air-water interface, we measured the desorption ~~of non adsorbed protein~~ by subphase exchange[16,18,19]. At pHs 5.5 and 9.0, the interfacial tension remained practically unchanged indicative of the stability of the adsorbed protein layer.  Close to the isoelectric point, pH 12.0, lysozyme appeared to be slightly desorbed (see Table 3), possibly due to its lower surface charge[35].

|  | ADS<br>γ (mN/m) | DES<br>γ (mN/m) | ADS<br>E (mN/m) |
|---|---|---|---|
| **Lysozyme pH 5.5** | 46.3±1.4 | 46.0±1.3 | 93±4 |
| **Lysozyme pH 9** | 48.7±1.6 | 46.6±1.8 | 84±11 |
| **Lysozyme pH 12** | 46.0±0.8 | 48.2±0.9 | 83±4 |
| **Pluronic (F68-O)** | 46±3 | 49±5 | 14±3 |
| **Pluronic (F68-W)** | 42.3±0.7 | 41.9±0.1 | 9.4±0.5 |

***Table 3.*** *Surface tension (**γ**) and dilatational modulus (**E**) (1 Hz) of lysozyme (4 mg/mL) at different pHs (5.5, 9.0 and 12.0), and Pluronic F68 (15 mg/mL) added in organic phase (F68-O), and in aqueous phase (F68-W), pH 5.5, after 50 min of adsorption (ADS) and desorption (DES). All the measurements were at room temperature.*

*Pluronic F68 at the air-water interface: adsorbed and spread layers*

Before addressing the interaction of lysozyme and Pluronic F68, we examined the properties of the Pluronic surface layer. We measured two types of surface layers: adsorbed from bulk aqueous solution (F68-W) and spread from organic solvent solution (F68-O). Table 3 shows the surface tension and dilatational modulus of these interfacial layers after 1 h. The possible desorption was also measured by reporting the surface tension after subphase exchange with pure buffer solution (1 h). These measurements were made in MilliQ water (pH5.5). The F68-W surface layer attained a lower surface tension and a lower dilatational modulus than the F68-O surface layer. In both cases, the surface layer showed negligible desorption and seemed very stable (Table 3). Our results agree with Torcello-Gómez et al., [36]. F68-W surface layers appeared loose and deformable whereas F68-O surface layer was slightly more elastic, suggesting that the organic phase favors the connectivity of the surface molecules. Actually, Pluronic F68 has a different conformation dissolved in water or chloroform owing to its triblock copolymer structure. In particular, F68 presents two large and hydrophilic tails, which extend in aqueous solution but fold in organic solution. This different conformation of Pluronic F68 in aqueous or organic phase could originate the different behavior of this molecule at the A/W interface.

*Lysozyme and Pluronic F68 at the air-water interface*

We now look into the interaction of lysozyme adsorbed layer with F68-W and with F68-O, hence mimicking the two different nanoparticle-synthesis procedures described above

Thus, we designed two different assays and made the measurements at three different pHs (5.5, 9.0 and 12.0) using the OCTOPUS device. Hence, onto a previously lysozyme adsorbed layer, we added Pluronic from the aqueous phase by subphase exchange of a F68 aqueous solution (F68-W) or we added the Pluronic from the organic phase by spreading a F68 organic solution on the surface layer (F68-O). Figures 3 and 4 show the surface tension/dilatational modulus of the surface layer during the sequential adsorption/desorption experiment, with the F68-W (Figures 3A and 4A) and F68-O (Figures 3B and 4B) procedures.

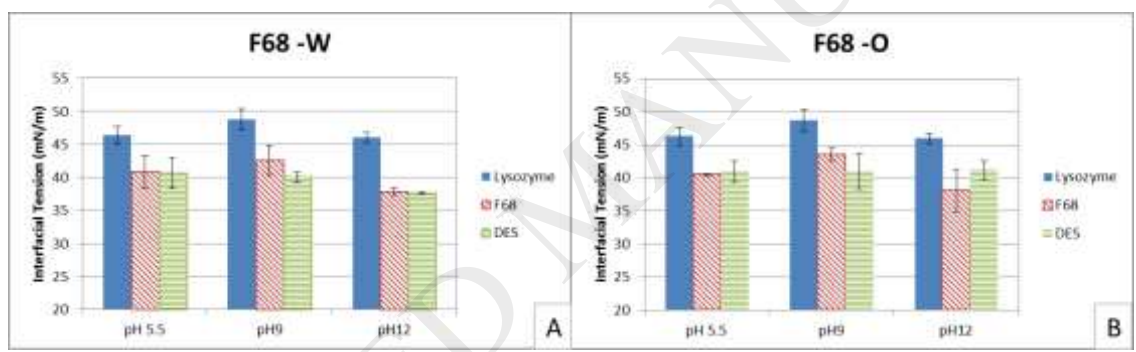

**Figure 3:** *Interfacial tension of adsorbed lysozyme (4 mg/ml) monolayer at air water interface, before (blue) and after (red) addition of Pluronic F68, and after the desorption experiment (green) at different pHs (5.5, 9.0, 12.0).* **A)** *Pluronic F68 (15 mg/ml) was added in the aqueous phase.* **B)** *Pluronic F68 (15 mg/ml) was added in the organic phase. All the measurements were made at room temperature.*

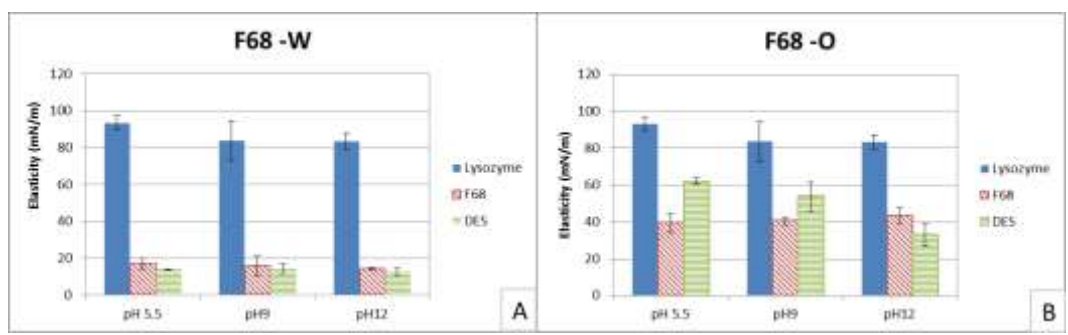

**Figure 4**: *Elasticity of lysozyme (4 mg/ml) monolayer at air/water interface, before (blue) and after (red) addition of Pluronic F68, and after the desorption experiment (green) at different pHs (5.5, 9.0, 12.0).* **A**) *Pluronic F68 (15 mg/ml) was added in aqueous phase.* **B**) *Pluronic F68 (15 mg/ml) was added in organic phase. All the measurements were made at room temperature.*

Let us first analyze the surface tension (Figure 3). The first step in both assays was identical, i.e. adsorption of lysozyme for 1 h. The results of this first step are hence the surface layers described in the previous section for lysozyme.

In the second step, we added the Pluronic F68. In both methods, and for all the different pHs measured, interfacial tension diminished (see Figure 3). Consequently, we assume that it occupies the air/water interface, forming a mixed surface layer with lysozyme. At pH 5.5, the addition of F68-W and F68-O onto the lysozyme monolayer lowered the surface tension to a similar value; $40.8 \pm 2.4$ mN/m and $40.5 \pm 0.2$ mN/m, respectively. This value differs from that of pure F68-W and F68-O as displayed in Table 3. This diminishing tendency suggests the formation of a mixed lys-F68 monolayer at the surface in both methods. At pH 9.0, the addition of Pluronic F68 onto the lysozyme monolayer decreased the surface tension, but to a lesser extent. Still, the value remained below that of pure lysozyme system, again accounting for the formation of a mixed monolayer but rather similar to that of the pure F68-W (Table 3). Finally, at pH 12.0, addition of F68-W and F68-O into the lysozyme monolayer decreased more notably the surface tension. The surface tension of the mixed monolayer fell to less than 40 mN/m, which is well below the surface tension of the pure systems (Table 3). This decrease possibly indicates an improved packing of the mixed monolayer achieved at pH 12.0. In summary, the way of adding the F68, aqueous or organic phase, does not appear to affect the final surface tension, but the pH of the lysozyme solution does. Close to the isoelectric point, the monolayer showed improved packing, consistent with the reduction of electrical repulsion. More information on the composition of this mixed interface is provided by the analysis of the dilatational modulus of these mixed layers in Figure 4. However, let us analyze first the surface tension of the third step of the sequential experiment (Figure 3).

In the last step, we studied the possible desorption (DES) of soluble material from the interface upon depletion of the bulk   solution by subphase exchange with the corresponding buffer (Figure 3). At pH 5.5, the monolayer was very stable, and the subphase exchange provided no change on the surface tension in any case. At pH 9.0, the surface tension of the monolayer decreased in both cases. This indicated that depletion of bulk material improved the packing of the lysozyme-F68 monolayers. Finally, at pH 12.0, the addition procedure of F68 on the lysozyme monolayer seemed to make a difference. While the lysozyme-F68-W monolayer remained unchanged after subphase exchange, the surface tension of the lysozyme-F68-O monolayer increased. These experimental results suggest that some adsorbed material was solubilized into the bulk phase when F68-O method was used, in contrast to the F68-W case.

Let us now evaluate the surface dilatational modulus of the sequential experiment shown in Figures 4 A and B. The dilatational modulus of the adsorbed layer of lysozyme at the air/water interface for different pHs was explained in the previous section. It should be noted here that, contrary to the surface tension, the dilatational modulus of the pure Pluronic F68-W and F68-O layer  significantly differed from that of lysozyme (Table 3). Hence, adsorbed lysozyme displayed high dilatational moduli at all pHs, while Pluronic F68 displays a low dilatational moduli, showing the lowest value for the F68-W method (Table 3).

The effect of adding Pluronic F68-W and F68-O onto the lysozyme-adsorbed layer at different pHs showed clear differences between the composition of the mixed layers. Figure 4 shows that the dilatational modulus was sharply reduced by the addition of F68 regardless of the pH and method. However, this reduction was clearly more notable for F68-W (Figure 4). These results indicate that the addition of F68-W resulted in a surface layer that was composed primarily of Pluronic, which seemed to displace the protein from the surface [37]. Previous studies have examined rheology at the air-water interface of mixed (protein-surfactant) adsorbed layers and the displacement of proteins by conventional surfactants (i.e. milk proteins) [38,39]. The coadsorption of low-molecular-weight surfactants onto the protein interfacial layer reduces the viscoelasticity and fluidizes this layer. The surfactants displace the proteins in the following way: firstly, the surfactant is adsorbed onto the protein layer, then compresses the protein network, and finally displaces the protein from the interface [38,39]. Blomqvist et al. showed that the triblock copolymer F127 was able to disrupt an adsorbed layer of βlactoglobulin and displace the protein from the surface [37]. Conversely, addition of F68-O appears to provide a mixed layer where the two species coexist at the surface, forming a medium cohesive network.

Also, we analyzed the dilatational modulus of the desorption step. On the one hand, the results from the F68-W method showed that the mixed layers, for all three pHs, were stable and did not alter their composition upon depletion of bulk material (see Figure 4). This agrees with the surface-tension results in Figure 3. On the other hand, the desorption profile for the F68-O method depended on the pH, in correlation with surface-tension (Figure 3). At pH 5.5 and 9.0, the surface dilatational moduli increased enhancing the cohesivity of mixed lysozyme-F68-O network. This evidences a major presence of lysozyme within the surface layer in agreement with the surface tension (Figure 3). Conversely, the surface dilatational modulus of the lysozyme-F68-O monolayer at pH 12.0 decreased to $33 \pm 6$ mN/m in the desorption step (Figure 4B), evidencing a major presence of pluronic.

In summary, the interfacial analysis shows that a mixed Pluronic-lysozyme layer forms at the interface irrespective of the method used for adding the surfactant. The F68-W method led to a stable layer composed mainly of Pluronic, which displaced the lysozyme towards the bulk. Since the layer is composed mainly of non ionic surfactant molecules, the pH of the lysozyme solution did not significantly affect this result. The F68-W method led to a stable layer with higher content of lysozyme and hence dependant on pH. Close to the isoelectric point (at pH 12.0), improved packing of the monolayer was evident, this being consistent with the reduction of electrical repulsion among protein molecules.

These interfacial results agree with the properties of the NPs synthesized with the counterpart methods. The F68-O method led to stable NPs only when the lysozyme solution was at pH 12.0, consistent with the stability of the interfacial layer being highest at this pH. All the F68-W-Lys NPs were stable over the short term, in agreement with a steric interaction, and for one month under storage at 4ºC for the F68-W-Lys9 and F68-W-Lys12 systems. The reason for not noting this pH effect on the interfacial results is that the PLGA polymer was not present in the interfacial study. Thus, the interaction lysozyme-PLGA must have been responsible for the pH effect on the F68-W-Lys NPs. When comparing F68-O-Lys12 and F68-W-Lys12 NPs, we found that the charge was less negative for the F68-W method. The interfacial results support the hypothesis of having more Pluronic F68 (a non-ionic surfactant) at the surface for F68-W-Lys12 NPs.

Additionally, the displacement of the lysozyme by the Pluronic provides an improved protection of the protein in the F68-W method. This importantly agrees with the better biological activity and better release pattern of the F68-W-Lys NPs that was found in a previous work [10].

3.  **CONCLUSIONS**

We present a new model to mimic *in vitro* the O/W interface in the first emulsion of the W/O/W synthesis method of NP, which enabled us to analyze the surfactant-protein (Pluronic F68-lysozyme) interaction at that interface.

The interfacial results showed that the procedure used to include the surfactant in the synthesis (through the oil (O) or the water phase (W)), and the conditions of the protein solution (pH) affected the interfacial conformation of the mixed surfactant-protein layer formed.

Addition of F68-W to lysozyme provides a stable surface layer composed mainly of Pluronic which is not affected by pH changes. Addition of F68-O to lysozyme provided a mixed surface layer with major presence of lysozyme at the surface hence, strongly dependant of pH. This is consistent with the colloidal properties of the nanoparticles resulting with the counterpart method. Moreover, F68-W displaced the lysozyme from the interface protecting its integrity and preserving its biological activity. These results very importantly agree with previous works showing a better biological activity and better release pattern of the F68-W-Lys NPs compared to the F68-O-Lys NPs [9].

In conclusion, the new *in vitro* model developed in this work allows evaluating the impact of the procedure chosen to include the surfactant in the NP synthesis (through the apolar or the polar phase). The results demonstrate that this choice strongly affects the protein/surfactant interaction, determining the biological activity of the encapsulated biomolecule.

**DECLARATION OF INTEREST**

Financial support granted by the following research projects: MAT2013-43922-R – European FEDER support included–(MICINN, Spain), RYC-2012-10556, MAT2015-63644-C2-1-R and PI12/2956.

Morris, Growth of surfactant domains in protein films, Langmuir. 19 (2003) 6032–6038. doi:10.1021/la034409o.